\def\year{2018}\relax

\documentclass[letterpaper]{article} 
\usepackage{aaai19}  
\usepackage{times}  
\usepackage{helvet}  
\usepackage{courier}  
\usepackage{url}  
\usepackage{graphicx}  
\usepackage{comment}
\usepackage{paralist}
\usepackage{booktabs}
\usepackage{subfig}
\usepackage{amsmath}
\usepackage{color}
\usepackage{colortbl}

\begin{document}
\title{What Propels Celebrity Follower Counts? Language Use or Social Connectivity}

\author{$^1$Jasabanta Patro, $^2$Rameshwar Bhaskaran, $^3$Animesh Mukherjee \\
$^{1,2,3}$IIT Kharagpur,\\
\{$^1$jasabantapatro, $^2$rameshwar.cs\}@iitkgp.ac.in,\\ 
$^3$animeshm@cse.iitkgp.ernet.in\\
}

\maketitle
\begin{abstract}
Follower count is a factor that quantifies the popularity of celebrities. It is a reflection of their power, prestige and overall social reach. In this paper we investigate whether the social connectivity or the language choice is more correlated to the future follower count of a celebrity. We collect data about tweets, retweets and mentions of 471 Indian celebrities with verified Twitter accounts. We build two novel networks to approximate social connectivity of the celebrities. We study various structural properties of these two networks and observe their correlations with future follower counts. In parallel, we analyze the linguistic structure of the tweets (LIWC features, syntax and sentiment features and style and readability features) and observe the correlations of each of these with the future follower count of a celebrity. As a final step we use there features to classify a celebrity in a specific bucket of future follower count (\textit{HIGH}, \textit{MID} or \textit{LOW}). We observe that the network features alone achieve an accuracy of 0.52 while the linguistic features alone achieve an accuracy of 0.69 grossly outperforming the network features. The network and linguistic features in conjunction produce an accuracy of 0.76. We also discuss some final insights that we obtain from further data analysis -- celebrities with larger follower counts post tweets that have (i) more words from `friend' and `family' LIWC categories, (ii) more positive sentiment laden words, (iii) have better language constructs and are (iv) more readable.

\end{abstract}
\section{Introduction}

The number of followers (aka follower count) that an individual has on a social media platform (e.g., Twitter) has become a symbol of `popularity', `prestige', `power' and is an indicator of the overall \textit{social reach} of the individual. While some debate, that this is only a `game'\footnote{https://techcrunch.com/2009/04/16/should-twitter-remove-its-follower-count/?guccounter=1}, there is a growing consensus that this is a determinant of social status and can also have monetary implications. In fact, various political, business and competition campaigns are reported to buy followers to propagate and make such campaigns successful\footnote{https://moz.com/blog/guide-to-buying-legit-twitter-followers, http://twitterboost.co/why-politicians-buy-twitter-followers-and-retweets/}.

Celebrities are no exceptions in this race of acquiring follower counts. In fact, they are, in most cases at the forefront of the race. There are certain celebrities who are known to have millions of followers and follower losses due to Twitter's policy change or otherwise makes a big news these days\footnote{https://www.teenvogue.com/story/celebs-lose-millions-of-twitter-followers-following-new-account-policy}. 

A pertinent question that arises is what strategies do celebrities employ early on to enhance their follower counts? Do they invest more on enriching their social connectivity or is the type/linguistic structure of their tweets that plays the key role in this enterprise. In the current paper we put forward for the first time this question and investigate in detail the correlations between social connectivity, language use and the follower count of 471 Indian celebrities with verified Twitter accounts.

Some of the key contributions of this work are

\noindent\textbf{Contributions}:
\begin{compactitem}
\item We prepare a list of 471 Indian celebrities with verified Twitter accounts. We collect all their tweets, their retweet and mention history as well as their follower counts.
\item We define novel types of retweet and mention networks and investigate the centrality properties of the nodes in these networks.
\item We perform an extensive linguistic analysis of the tweet text for each individual celebrity. In particular, we extract LIWC features, syntactic features, sentiment features, style features as well as readability features. 
\item Finally, we build a classifier to predict the range of future follower count of a celebrity using the network and the language features that we extract.
\end{compactitem}

\noindent\textbf{Key results and observations}:
Some of the important results and observations are as follows,

\begin{compactitem}
\item We observe that network features such as betweenness, degree and PageRank centrality are positively correlated to the future follower count. In contrast, clustering coefficient is negatively correlated to the future follower count. 
\item We observe that certain LIWC features such as `positive emotion', `affection', `cognitive mechanics' and `social' show high correlation with the future follower count; in contrast, features like `sad', `anger', `anxiety', `death' and `swear words' show low correlation with the future follower count.
\item The classifiers trained using only network features achieve an accuracy of 0.52 in predicting the future follower count. On the other hand, the classifiers trained using linguistic features achieve an accuracy of 0.69 which by far outperforms the network features. The combination of both the network and the linguistic features results in an overall accuracy of 0.76 in predicting the future follower count.   
\item Some of the interesting observations are -- (i) LIWC categories like `anger' and `negate' are in larger proportions for celebrities who are less popular (i.e., lower follower counts). On the other hand, highly popular celebrities (i.e., higher follower counts) have larger proportions of words from the LIWC categories like `family' and `friend' in their tweets, (ii) tweets of celebrities with higher follower counts are laden with positive sentiments and (iii) celebrities with larger follower counts post tweets that seem to have better language construct and are more readable. 
\end{compactitem}
\section{Related works}
\noindent\textbf{Celebrities in news}: Castillo et al.~\cite{Castillo2013SaysWA} analyzed linguistic style of people when they post news related to the celebrities. Sweetser et al.~\cite{Sweetser2008StealthSP} studied the effects of personalized and `stealth' political discourse on the weblogs (or blogs), and repercussions on the levels of political trust, information efficacy and political uses/gratifications. Ogan et al.~\cite{Ogan2006ConfessionRA} in their study found that diversion drives most reading on the site, but social interaction provides the largest gratification to those who participate through writing confessions, commenting on others’ confessions and meeting people offline. Moynihan et al.~\cite{Moynihan2004TheIM} studied effect of celebrity marketing in pharmaceutical domain. Sheridan et al.~\cite{Sheridan2007CelebrityWA} studied the concept of celebrity worship and its relation to criminality and addiction. Cheng et al.~\cite{Cheng2007TheIO} studied the influence of media reporting of the celebrity suicides  on the suicide rates. Hayward et al.~\cite{Hayward2004BelievingOO} studied the causes and consequences of a CEO Celebrity. Maltby et al.~\cite{Maltby2004PersonalityAC} studied the relation between celebrity worship and mental health. Elberse et al.~\cite{Elberse2011TheEV} studied economic value of celebrity endorsements. 

\noindent\textbf{Celebrities on social media}: There have been a number of studies investigating the social behavior of celebrities. Kumar et al.~\cite{Kumar2015DetectingCI} proposed a topic model analysis of social media content, following celebrity suicides which revealed the presence of derogatory tone in the content. Romero et al.~\cite{Romero2011InfluenceAP} proposed an algorithm to measure the influence and passivity of users, based on information forwarding activity. Cha et al.~\cite{Cha2010MeasuringUI} did a comparative study of user influence across topics with three influence measures: in-degree, retweets and mentions. Sakaki et al.~\cite{Sakaki2010HowTB} identified various parameters related to social networks that are found to be a factor for celebrity popularity. Marwick in his study \cite{Marwick2011ToSA} described various celebrity practices including language of acknowledging fans and cultural references to create fan affiliation. Hoffman et al.~\cite{Hoffman2015BiologicalPA} described biological, psychological and social processes that explain celebrities' influence on patients' health-related behaviors. Zhao et al.~\cite{Zhao2014ACA} proposed a computational approach to measure the correlation between expertise and social media influence, for celebrities on microblogs. Kim et al.~\cite{Kim2014FindingIN} concluded that, in social networks, information can be efficiently propagated using neighbors having high potential of propagation rather than having high number of neighbors. Brzozowski et al.~\cite{Brzozowski2011WhoSI} in their study compared a variety of features for recommending users, and presented design implications for social networking services. Bakshy et al.~\cite{Bakshy2011EveryonesAI} proposed several measures to quantify influence in the social networks like Twitter. Sharma et al.~\cite{Sharma2014InferringST} proposed methods to infer social ties from common activities in Twitter. Abbasi et al.~\cite{Abbasi2014AmIM} analyzed homophily effect in the directed social networks. Taxidou et al.~\cite{Taxidou2015ModelingID} modeled information diffusion in the social media. Zhou et al.~\cite{Zhou2016TweetPA} analyzed the deleted tweets to understand and identify regrettable ones. Taxidou et al.~\cite{Taxidou2014OnlineAO} studied effect of different influence models on the cascades.

\noindent\textbf{Present work}: Our work is different from the ones reviewed above. We study the celebrity profiles, their social connectivity and tweeting behavior and observe correlations of these variables with the future follower count. We show for the first time that the language use is way more important than social connectivity in acquiring follower counts. 
\section{Dataset description}

\subsection{Data collection process}

\begin{figure}[!htbp]
\centering
 \includegraphics[width=50mm]{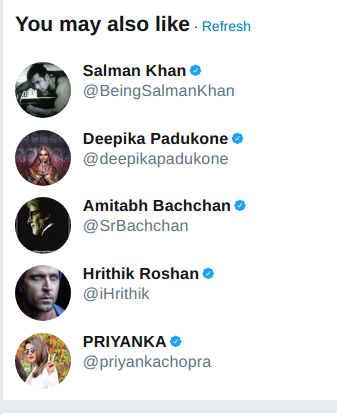}
 \caption{List of celebrities in the ``You may also like'' section of the Indian celebrity Shah Rukh Khan's profile.}
 \label{Figure:twitter_recommendation}
\end{figure}

In this study, we consider a list of 471 Indian celebrities\footnote{Celebrity list: \url{https://github.com/zorroblue/language-matters/tree/master/data}} from five different areas -- \textit{Movie, Music, News, Tech and Sports}. We start with certain verified celebrity accounts and then use Twitter's recommendations in the ``You may also like'' section (Figure~\ref{Figure:twitter_recommendation}) to get more celebrities. In most cases, these recommendations correspond to colleagues of the initial set of celebrities in the same category. Table~\ref{Table:CelebDist} shows the distribution of the number of celebrities across the different categories.

\begin{table}[!htbp]
\centering
\caption{Distribution of celebrities across the five categories.}
\label{Table:CelebDist}
\begin{tabular}{|l|c|}
\hline
Category & $|celebrities|$ \\
\hline
Movies & 92 \\
Music & 95 \\
News & 92 \\
Tech & 95 \\
Sports & 97 \\\hline
Total & 471
\\\hline 
\end{tabular}
\end{table}

We gathered tweets generated by these handles using the Twitter streaming API\footnote{Twitter streaming API: \url{https://developer.twitter.com/en/docs/tutorials/consuming-streaming-data.html}} for the duration of  June and July 2017. We collected the tweets of two different categories -- (i) tweets posted by the celebrities themselves in this period, and (ii) tweets posted by other users who either mention or retweet one or more of these celebrities. We removed  all the invalid and duplicate tweets. We also collect the future follower count of each celebrity\footnote{Follower count list: \url{https://github.com/zorroblue/language-matters/tree/master/data}} for the month of October 2017. 

\subsection{Basic statistics of the data collected}
\begin{figure}[!htbp]
\centering
 \includegraphics[width=75mm]{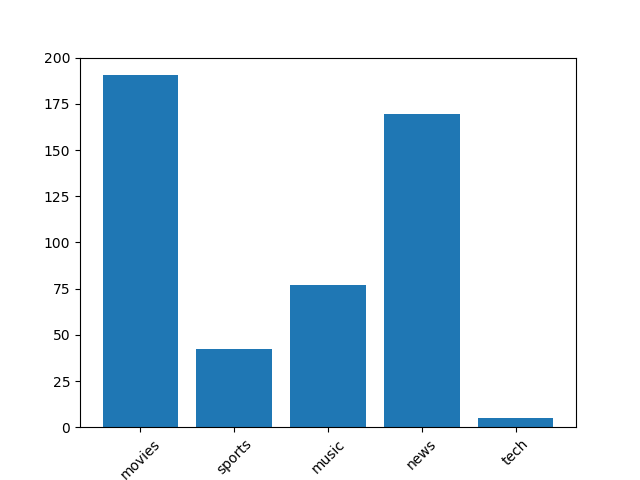}
 \caption{Average retweet density vs category}
 \label{Figure:CelebrityAverageRetweetDensity}
\end{figure}

The data collection and filtering process resulted in 23,57,070 tweets, out of which 15269 are tweets from the celebrities. We calculate \textit{average retweet density} ($ARD$) for each category of celebrities by taking the ratio of the cumulative count of the number of retweets obtained by the celebrities' tweets to the total number of tweets done by the celebrities in that category. Figure~\ref{Figure:CelebrityAverageRetweetDensity} shows the $ARD$ across the five categories. Interestingly, the \textit{movie} and the \textit{news} categories have the highest $ARD$, the former possibly due to the huge fan following of movie stars and the latter due to the sharing of the latest and `hot' news.


\section{Social connectivity}
In this section we construct two different networks that approximate the social connectivity of the celebrities. One of these is based on retweets and the other on mentions. We finally extract various features from these networks and observe how they correlate with the celebrity follower buckets. 


\subsection{Retweet network}
We consider each celebrity as a node in this network.  There is an edge between two nodes if at least five common users\footnote{The number five has be set empirically.} have retweeted both of their tweets (may be different tweets) in our dataset. The edge weight is the normalized number of common retweeters, given by
\begin{equation}
weight_{rt}(A, B) = \frac{common\_retweeters(A, B)} {\sum_{(i,j)} common\_retweeters(i, j)} 
\end{equation}
Using this criteria, we get a network of 324 nodes and 20502 edges. 

\subsection{Mention network} Similar to the retweet network, here we consider celebrities as nodes. However, here an edge between two nodes form if at least five common users\footnote{The number five has be set empirically.} have mentioned both celebrities independently in their tweets in our data set. The edge weight is the normalized number of common mentioners, given by 
\begin{equation}
weight_{men}(A, B) = \frac{common\_mentioners(A, B)} {\sum_{(i,j)} common\_mentioners(i, j)} 
\end{equation} 

Using this criteria we obtain 368 nodes and 44072 edges.  

\subsection{Network features} 
We extract the following network properties from the retweet as well as the mention network.

\subsubsection{Network centrality measures} We compute traditional network centrality measures such as betweenness centrality ($C_{bet}$)~\cite{freeman1977set}, closeness centrality ($C_{clo}$)~\cite{sabidussi1966centrality}, clustering coefficient ($Clust_{coff}$)~\cite{holland1971transitivity}, degree centrality ($C_{deg}$)~\cite{mej2010networks} and PageRank centrality ($C_{pr}$)~\cite{sullivan2007google} \if{0}eigenvector centrality ($C_{eig}$)~\cite{newman2003structure}\fi for our analysis. We obtain the Spearman's rank correlation between the follower count and these network measures for all the celebrities. In the Table~\ref{tab:netfet} we report these correlations. We observe that in both the networks, betweeness, degree and PageRank centralities are strongly positively correlated to the future follower count. On the other hand, cluster coefficient is negatively correlated to the future follower count in both the networks.


\begin{table}
\centering
\caption{Spearman's rank correlation between network features and the follower counts.}
\label{tab:netfet}
\begin{tabular}{|l|l|l|}
\hline
Feature & $\rho_{rt}$ & $\rho_{men}$ \\
\hline
$C_{clo}$ & 0.14	 & -0.13   \\
$C_{bet}$ & 0.55	 & 0.57    \\
$C_{deg}$ &	0.63 	 &  0.59  \\
$Clust_{coff}$ & -0.43  & -0.57  \\
$C_{pr}$ & 0.57 & 0.58 
\\\hline 
\end{tabular}
\end{table}

\section{Linguistic structure of the tweets}

In this section we analyze the linguistic structure of the tweets posted by the celebrities. Prior to the analysis, we preprocess all tweets by removing non-ASCII characters, urls, ellipses, special characters like \#, @, \{, \} and stop words, followed by word stemming using the Porter stemmer\footnote{\url{https://tartarus.org/martin/PorterStemmer/index.html}}. 

\subsection{LIWC analysis}
As a first step, we compute the fraction of different LIWC\footnote{LIWC Companion :\url{http://www.liwc.net/comparison.php}} categories in the celebrity tweets. \if{0}The LIWC dictionary has 64 predefined categories as shown in Table \ref{Table:LIWCCat}.\fi For every individual celebrity, we compute the fraction of words per tweet in each LIWC category. We term this fraction as the category density. Based on this factor we prepare a rank list of celebrities for each category. We then compute the Spearman's rank correlation between these rank lists and the follower count based ranks. Tables~\ref{Table:TopLIWCSpearman} and~\ref{Table:BottomLIWCSpearman} show the top and the bottom ten LIWC categories that have the largest and the smallest correlations. We observe that the categories like `positive emotion', `affect', `cognitive mechanism' and `social' are highly correlated to future follower count. Categories like `assent', `death' and `swear words' are least correlated to future follower count.

\if{0}
\begin{table}[!htbp]
\centering
\caption{LIWC categories}
\label{Table:LIWCCat}
\begin{tabular}{|l|l|l|l|}

\hline
Achiev   &  Adverbs   &  Affect   &  Anger   \\  
Anx      &  Article   &  Assent   &  AuxVb   \\ 
Bio      &  Body      &  Cause    &  Certain \\  
CogMech  &  Conj      &  Death    &  Discrep \\ 
Excl     &  Family    &  Feel     &  Filler  \\  
Friends  &  Funct     &  Future   &  Health  \\ 
Hear     &  Home      &  Humans   &  I       \\  
Incl     &  Ingest    &  Inhib    &  Insight \\ 
Ipron    &  Leisure   &  Money    &  Motion  \\  
Negate   &  Negemo    &  Nonflu   &  Numbers \\ 
Past     &  Percept   &  Posemo   &  Ppron   \\
Prep     &  Present   &  Pronoun  &  Quant   \\ 
Relativ  &  Relig     &  Sad      &  See     \\
Sexual   &  SheHe     &  Social   &  Space   \\ 
Swear    &  Tentat    &  They     &  Time    \\
Verbs   &  We   &  Work   &  You 
\\\hline 
\end{tabular}
\end{table}
\fi

\begin{table}[h]
\centering
\caption{Top ten LIWC categories showing higher Spearman's correlation with follower count based ranking.}
\label{Table:TopLIWCSpearman}
\begin{tabular}{|l|l|}
\hline
LIWC category & $\rho$ \\
\hline
Posemo &	0.71  \\
Affect &	0.70  \\
Funct &		0.68  \\
CogMech &	0.67  \\
Social &	0.66  \\
Relativ &	0.66  \\
Article &	0.66  \\
Prep &		0.65  \\
Pronoun &	0.65  \\
Incl &		0.64 
\\\hline 
\end{tabular}
\end{table}

\begin{table}[!htbp]
\centering
\caption{Bottom ten LIWC categories showing lower Spearman's correlation with follower count based ranking.}
\label{Table:BottomLIWCSpearman}
\begin{tabular}{|l|l|l|l|}
\hline
LIWC category & $\rho$ \\
\hline
SheHe &		0.31 \\
Sad & 		0.28 \\
Anger & 	0.27 \\
Filler & 	0.27 \\
Nonflu & 	0.24 \\
Ingest & 	0.24 \\
Anx & 		0.23 \\
Assent & 	0.23 \\
Death & 	0.19 \\
Swear & 	0.11 
\\\hline 
\end{tabular}
\end{table}

\subsection{Use of in-vocabulary words}

In this section we analyze the propensity of the use of in-vocabulary words by the different celebrities. For this purpose, we compute the ratio of the total number of in-vocabulary to the out-of-vocabulary words from all the tweets posted by each celebrity. We use the GNU Aspell dictionary\footnote{\url{http://aspell.net/}} to find the number of in-vocabulary words. We then rank the celebrities based on this ratio and report Spearman's rank correlation with the follower counts. We observe very low negative rank correlation with the value -0.057.

\subsection{Tweet sentiment analysis}

In this section we analyze the overall sentiment in the tweets posted by each celebrity. We use the NLTK's VADER sentiment extraction tool\footnote{\url{http://www.nltk.org/_modules/nltk/sentiment/vader.html}} for our analysis. The analyzer returns four different scores -- positive ($pos$), negative ($neg$), neutral ($neu$) and compound ($comp$). The last score is calculated by applying a normalized function over the first three scores. We rank the celebrities using the above scores and compute the Spearman's rank correlation with the follower counts. The results are reported in Table~\ref{Table:Sentiment}. While positive sentiment is correlated with future follower count the negative sentiment is anti-correlated. Highly followed celebrities therefore seem to have more positive sentiment in their tweets.

\begin{table}[!htbp]
\centering
\caption{Spearman's rank correlation between sentiment based rank lists and follower count based rank list.}
\label{Table:Sentiment}
\begin{tabular}{|l|l|l|l|}
\hline
Sentiment & $\rho$ \\
\hline
$pos$ &	0.17 \\
$neg$ & -0.12\\
$neu$ &	-0.05 \\
$comp$ & 0.21 
\\\hline 
\end{tabular}
\end{table}

\subsection{POS tag entropy analysis} 

For the sentences to be well formed, intuitively the probability distribution of POS tags in the sentences has to follow a uniform distribution~\cite{marquez2000machine}. This means that the following expression for entropy over the POS tag probability distribution has to be maximum for the most well formed sentences. 
\begin{equation}
Entropy_{pos} = -\sum_{X \epsilon pos\-tags} p(X)\log p(X)
\end{equation}

We rank the celebrities using this entropy value and compute the correlation between this rank list and the follower count based rank list. We obtain a low Spearman's rank correlation value of -0.04.

\subsection{Style feature analysis} 

Style is one of the key factors in any linguistic analysis.~\cite{Karlgren:1997:VSV:938438.938866} reported various measures of styles in running text (see Table~\ref{Table:StyleFeatures}). We compute these measures from the collection of tweets of each individual celebrity. In Table~\ref{tab93} we report Spearman's rank correlation between the style feature based rank list and the follower count based rank list of the celebrities. None of the style features seem to be strongly correlated to the future follower count.


\begin{table*}[!htbp]
\centering
\caption{Different style metrics discussed in \cite{Karlgren:1997:VSV:938438.938866}}
\label{Table:StyleFeatures}
\begin{tabular}{|l|l|l|}
\hline
Variable name & Statistic & Typical Range \\
\hline
$TTR$ & Type token ratio & 0.13-0.89 \\
$CPW$ & Average word length in characters & 4.59-9.95 \\
$WPS$ & Average sentence length in words & 2.45-63.1 \\
$P1$ & Proportion first person pronouns of words & 0-105 \\
$P2$ & Proportion second person pronouns of words & 0-20 \\
$P3$ & Proportion third person pronouns of words & 0-60 \\
$IT$ & Proportion 'it' of words & 0-44
\\\hline 
\end{tabular}
\end{table*}

\begin{table}[h]
\centering
\caption{Spearman's correlation between style based features' rank lists and follower count based rank list.}
\label{tab93}
\begin{tabular}{|l|l|}
\hline
Style feature & $\rho$ \\
\hline
$TTR$ & 0.03 \\
$CPW$ &-0.05 \\
$WPS$ & 0.01 \\
$P1$ & -0.08 \\
$P2$ & 0.04  \\
$P3$ & -0.10 \\
$IT$ & -0.07 
\\\hline 
\end{tabular}
\end{table}

\subsection{Readability analysis} 

Readability is a way to quantify the reading convenience of a running text. Usually estimations are done by counting the number of syllables, words and sentences. While there are quite a few quantitative variants, the automatic readability index ($ARI$) is the most popular one. The $ARI$ is defined as, 

\begin{equation}
ARI = 4.17  (\dfrac{characters}{words}) + 0.15 (\dfrac{words}{sentences}) - 21.53
\end{equation}

As per the above definition, the lower the value of $ARI$ the better. We rank the celebrities again by $ARI$ scores and compute the Spearman's rank correlations with the follower counts. Once again we obtain a low Spearman's rank correlation of -0.01.

\section{Popularity prediction}

In this section we predict the popularity, i.e., the follower count bucket for a celebrity, using network and linguistic features in turn. We present separate results of predictions using only network features as well as only linguistic features. We also study the effect of using different categories of linguistic features on the prediction results. As a final step, we combine the most relevant network and linguistic features for the purpose of prediction. We use supervised classification methods with ten fold cross-validation for the popularity prediction.

\subsection{Dataset for classification}

We place the celebrities into one of the three buckets -- \textit{HIGH}, \textit{MID} or \textit{LOW}. We consider only those celebrities who are present in both the retweet and the mention network. In effect, therefore we have 324 celebrities for classification. There are 108 celebrities in the \textit{HIGH} and the \textit{MID} bucket each in order of the number of their follower counts. The rest are placed in the \textit{LOW} bucket.

\subsection{Network features}

\subsubsection{All network features} Here, we consider all the network features extracted from the retweet and the mention network for the classification. The accuracy obtained for various classifiers \if{0}excluding and including community features (i.e. $Comm_{exclude}$ and $Comm_{include}$)\fi are shown in Table \ref{Table:AccuracyCommunityAll}. 

\begin{table}
\centering
\caption{Accuracy of classifiers considering all the network features.}
\label{Table:AccuracyCommunityAll}
\begin{tabular}{|l|l|}
\hline
Classifier & Accuracy\\ 
\hline
Random forest & 0.51\\ 
XGBoost & 0.51\\ 
SGD Classifier  & 0.49 \\ 
\rowcolor{green} Guassian Naive Bayes  & \textbf{0.52} 
\\\hline 
\end{tabular}
\end{table}

\subsubsection{Few highly correlating network features} 
Here we experiment with various highly correlating network features to predict the class. We observe that the following features -- betweenness, degree and PageRank centralities, and the clustering coefficient of both the retweet and the mention network work as the best combination of features. 

The accuracy of the different classifiers for these set of features are noted in Table \ref{Table:Fewnetwork}. 

\begin{table}
\centering
\caption{Accuracy of classifiers considering few highly correlating network features.}
\label{Table:Fewnetwork}
\begin{tabular}{|l|l|}
\hline
Classifier & Accuracy \\
\hline
Random forest & 0.50 \\
\rowcolor{green}XGBoost & \textbf{0.54} \\
SGD Classifier  & 0.40 \\
Guassian Naive Bayes  & 0.50
\\\hline 
\end{tabular}
\end{table}

\subsection{Linguistic features}

\subsubsection{All linguistic features} Here we consider all the linguistic features as described in the previous section for the classification. The accuracy obtained from the different classifiers are noted in Table~\ref{Table:AllLingAcc}. We observe that the linguistic features by far outperforms the network features in classification. This indicates that the choice of language plays a very crucial role in framing the future popularity of a celebrity. In fact, this choice is much more important than building social connections.

\begin{table}
\centering
\caption{Classification accuracy using all the linguistic features.}
\label{Table:AllLingAcc}
\begin{tabular}{|l|l|l|l|l|l|l|l|}
\hline
Classifier & Accuracy \\
\hline
\rowcolor{green}Random forest & \textbf{0.69} \\
XGBoost & 0.67 \\
SGD Classifier & 0.34 \\
Gaussian Naive Bayes & 0.63 
\\\hline 
\end{tabular}
\end{table}

\subsubsection{Only LIWC features} Here we consider all the LIWC features for the classification. The results are shown in Table~\ref{Table:LIWCLingAcc}. The results show that among the linguistic features, the LIWC features themselves are one of the strongest discriminators.  

\begin{table}
\centering
\caption{Classification accuracy using all the LIWC features.}
\label{Table:LIWCLingAcc}
\begin{tabular}{|l|l|}
\hline
Classifier & Accuracy \\
\hline
\rowcolor{green}Random forest & \textbf{0.66} \\
XGBoost & 0.65 \\
SGD Classifier & 0.61 \\
Gaussian Naive Bayes & 0.63 
\\\hline 
\end{tabular}
\end{table}

\subsubsection{Linguistic features other than LIWC} The classification results for different classifiers considering all linguistic features except the LIWC categories are shown in Table~\ref{Table:ExclLIWCLingAcc}. In isolation, these features do not seem to perform well. 

\begin{table}
\centering
\caption{Classification accuracy for all the linguistic features excluding LIWC.}
\label{Table:ExclLIWCLingAcc}
\begin{tabular}{|l|l|}
\hline
Classifier & Accuracy \\
\hline
Random forest & 0.34 \\
\rowcolor{green}XGBoost & \textbf{0.38} \\
SGD Classifier & 0.30 \\
Gaussian Naive Bayes & 0.35 
\\\hline 
\end{tabular}
\end{table}

\subsubsection{Few handpicked linguistic features} Here we consider only those linguistic features that show high Spearman's rank correlation with follower count. These features include positive emotion (`posemo'), affection words (`affect'), function words (`funct'), cognitive words (`cogmech') and social words (`social') from LIWC. The accuracy of classifiers using these features are shown in Table~\ref{Table:FewLingAcc}. It turns out that using all linguistic features marginally improves the accuracy over using these set of handpicked features. 

\begin{table}
\centering
\caption{Classification accuracy using highly correlating linguistic features.}
\label{Table:FewLingAcc}
\begin{tabular}{|l|l|l|l|l|l|l|l|}
\hline
Classifier & Accuracy \\
\hline
\rowcolor{green}Random forest & \textbf{0.66} \\
XGBoost & 0.65 \\ 
SGD Classifier & 0.6 \\
Gaussian Naive Bayes & 0.656
\\\hline 
\end{tabular}
\end{table}

\subsection{Network + linguistic features:} Here we consider a subset of linguistic and network features showing high Spearman's rank correlation with follower count to predict the popularity of celebrities. The linguistic features in this subset include `affect' words, function words (`funct'), cognitive words (`cogmech') and `social' words from LIWC and compound sentiment from the tweet sentiment analysis. Similarly, the network features in this subset include betweenness, degree and PageRank centralities and the clustering coefficient of the retweet and mention network.

\begin{table}
\centering
\caption{Classification accuracy for a mix of linguistic and network features}
\label{tab91}
\begin{tabular}{|l|l|}
\hline
Classifier & Accuracy \\
\hline
Random forest & 0.73 \\
XGBoost & 0.71 \\ 
SGD Classifier & 0.46 \\ 
\rowcolor{green} Gaussian Naive Bayes & \textbf{0.76}
\\\hline 
\end{tabular}
\end{table}

\section{Discussion}
In this section we report some of the interesting insights that we find from the further analysis of the data. 

\noindent \textbf{Network centrality measures}: We compute and report the average centrality values for both the retweet and the mention network in Table~\ref{tab:bucknetcent}. The betweenness and the degree centralities in the retweet and the mention networks for the \textit{HIGH} bucket are drastically larger (3 to 4 times) than the other two buckets. 

\begin{table}[h]
\centering
\caption{ Bucket wise average network centrality measures. }
\label{tab:bucknetcent}
\begin{tabular}{|l|l|l|l|}
\hline
Feature & $HIGH$ & $MID$ & $LOW$ \\
\hline
$RT-C_{clo}$  & 0.086	 & 0.085   & 0.082 \\
$MEN-C_{clo}$ & 0.078	 & 0.08    & 0.08  \\
$RT-C_{bet}$  & 221.57	 & 52.05   & 36.09 \\
$MEN-C_{bet}$ & 140.35	 & 50.55   & 28.76   \\
$RT-C_{deg}$  &	186.81 	 & 120.43  & 72.42 \\
$MEN-C_{deg}$ &	312.12 	 & 263.71 &  214.94\\
$RT-Clust_{coff}$ & 0.89  & 0.94 & 0.91\\
$MEN-Clust_{coff}$ & 0.96 & 0.98 & 0.99 \\
$RT-C_{pr}$ & 0.005 & 0.002 & 0.001 \\
$MEN-C_{pr}$ & 0.005 & 0.002 & 0.001
\\\hline 
\end{tabular}
\end{table}

\noindent \textbf{LIWC}: Some of the interesting insights that we obtain from the LIWC analysis is illustrated in Figure~\ref{Figure:SelectedLIWCCats}. We note that while celebrities in the $HIGH$ bucket mostly tweet about `friend' and `family', those in the $LOW$ bucket tend to tweet about matters related to `anger' and `negation'. This further shows that how language choice could be a very crucial factor in framing the overall `social' reputation of a celebrity.        

\begin{center}
    \includegraphics[width=0.5\textwidth]{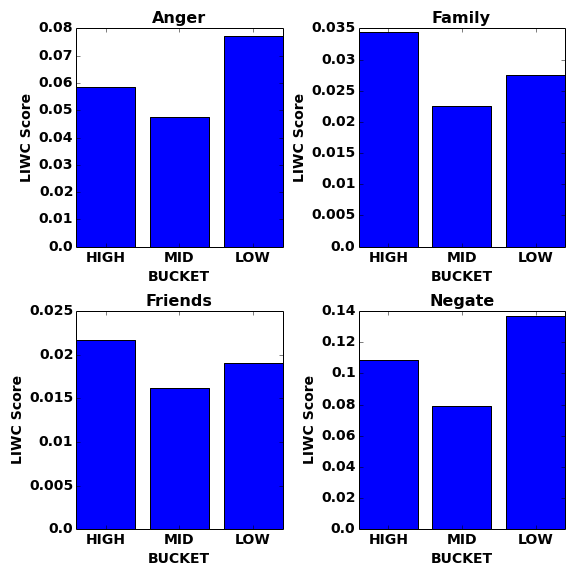}
    \captionof{figure}{Selected LIWC categories.}
 \label{Figure:SelectedLIWCCats}
\end{center}

\noindent \textbf{Proportion of in-vocabulary words}: We calculate average proportion of in-vocabulary words present in the tweets of celebrity belonging to each bucket. The results are shown in Table~\ref{table:InvocaProportion}. The \textit{HIGH} bucket as usual indicate higher proportion of use of in-vocabulary words. However, what is more interesting is the the \textit{LOW} bucket also indicate a similar extent of use of in-vocabulary words. On manual inspection we found that the \textit{LOW} bucket mostly comprises celebrities from the news category. Consequently, the usage of high proportions of in-vocabulary words is justified.  

\begin{table}[!htbp]
\centering
\caption{Bucket wise proportion of in-vocabulary words.}
\label{table:InvocaProportion}
\begin{tabular}{|l|l|l|}
\hline
$HIGH$ & $MID$ & $LOW$ \\
\hline
0.89 & 0.86 & 0.89
\\\hline 
\end{tabular}
\end{table}

\noindent \textbf{Tweet sentiments}: We present the average sentiment score for each of the three follower count based buckets in Table~\ref{Table:GroupSentiment}. Clearly, the \textit{HIGH} bucket has the tweets with the largest positive sentiment whereas the celebrities in the \textit{LOW} bucket have slightly high negative sentiment in their tweets.
\begin{table}[!htbp]
\centering
\caption{Bucket wise average score of tweet sentiments.}
\label{Table:GroupSentiment}
\begin{tabular}{|l|l|l|l|}
\hline
Category & \textit{HIGH} & \textit{MID} & \textit{LOW} \\
\hline
$pos$ & 0.26 & 0.19 & 0.18 \\
$neg$ & 0.04 & 0.05 & 0.06 \\
$neu$ & 0.61 & 0.65 & 0.68 \\
$comp$ & 0.25 & 0.17 & 0.14
\\\hline 
\end{tabular}
\end{table}

\noindent \textbf{POS tag entropy}: The average entropy values in each of the buckets are shown in Table~\ref{Table:GroupWiseEntropy}. The \textit{HIGH} bucket shows a slightly larger entropy (i.e., better language construct) compared to the other two buckets. 

\begin{table}[!htbp]
\centering
\caption{Bucket wise value of average POS tag entropy.}
\label{Table:GroupWiseEntropy}
\begin{tabular}{|l|l|l|}
\hline
$HIGH$ & $MID$ & $LOW$ \\
\hline
3.23 & 3.21 & 3.17
\\\hline 
\end{tabular}
\end{table}

\noindent \textbf{Style features}: We report the average values of various style features for each of the follower count buckets in Table~\ref{Table:GroupStyleFeature}. There is no significant difference observable among the three categories.

\begin{table}
\centering
\caption{Bucket wise average score of style features.}
\label{Table:GroupStyleFeature}
\begin{tabular}{|l|l|l|l|}
\hline
Category & $HIGH$ & $MID$ & $LOW$ \\
\hline
$TTR$ & 0.21 & 0.23 & 0.2 \\
$CPW$ & 4.29 & 4.39 & 4.39 \\
$WPS$ & 11 & 10 & 11 \\
$P1$ & 0.01 & 0.01 & 0.02 \\
$P2$ & 0.02 & 0.02 & 0.01 \\
$P3$ & 0.01 & 0.01 & 0.01 \\
$IT$ & 0.01 & 0.01 & 0.01
\\\hline 
\end{tabular}
\end{table}

\noindent \textbf{Readability}: We calculate the average $ARI$ value of the celebrities in the \textit{HIGH}, the \textit{MID} and the \textit{LOW} buckets (see Table~\ref{Table:AvgARI}). We clearly observe that the celebrities in the \textit{HIGH} bucket post the most readable tweets. 
 
\begin{table}
\centering
\caption{Bucket wise value of average $ARI$.}
\label{Table:AvgARI}
\begin{tabular}{|l|l|l|}
\hline
$HIGH$ & $MID$ & $LOW$ \\
\hline
4.23 & 4.25 & 4.76
\\\hline 
\end{tabular}
\end{table}

\section{Conclusion}

In this paper we have presented a verity of linguistic features over celebrities' tweets and network features over retweet and mention network to do an extensive analysis of the features correlating with popularity of celebrities. We have seen that a compact set of LIWC features including positive emotion, affection, function, cognitive and social words outperform highly correlating network features in predicting the popularity. Thus from this work we can conclude that language bears an important role in propelling popularity/follower count of celebrities. This indeed a strong message to aspiring candidates for celebrities to carefully choose their language over social media so that they can get more popularity. Finally, we have seen that best accuracy(0.76) is achieved for a combination of linguistic and network features. In future we would like to expand span of linguistic and network features to make an even more in-depth study of relation of these features with the popularity of celebrities.

\bibliographystyle{aaai}
\bibliography{sample-bibliography}

\end{document}